# Crack Formation in Ceramic Films Used in Solid Oxide Fuel Cells

Xin Wang, Zhangwei Chen and Alan Atkinson
Department of Materials, Imperial College London SW7 2BP, UK

**Abstract**

The manufacture of solid oxide fuel cells (SOFCs) involves fabrication of a multilayer ceramic structure, for which constrained sintering is a key processing step in many cases. Defects are often observed in the sintered structure, but their formation during sintering is not well understood. In this work, various ceramic films were fabricated by screen printing and a variety of defects observed. Some films showed "mud-cracking" defects, whereas others presented distributed large pores. "Mud cracking" defects were found to originate from a network of fine cracks present in the green film and formed during drying and binder burn-out. Control of these early stages is essential for producing crack-free films. In order to investigate how defects evolve during sintering, artificial cracks were introduced in the green films using indentation. It was observed that crack opening always increased during constrained sintering. In contrast, similar initial cracks could be closed and healed during co-sintering. **Published in *Journal of the European Ceramic Society 33 (13-14), 2539–2547***

## 1 Introduction

A fuel cell is an electrochemical device capable of converting chemical energy directly to electricity with high efficiency. SOFCs (Solid Oxide Fuel Cells) offer many advantages over other fuel cell types. SOFCs can be used with a variety of fuels from hydrogen to hydrocarbons with a minimum of fuel processing and may also play a role in carbon sequestration strategies [1, 2]. The basic structure of a SOFC comprises at least three layers of ceramics or cermet: i.e., electrolyte, anode and cathode layers. The electrolyte layer must be a dense ceramic, and typical materials are yttria stabilised zirconia (YSZ) or gadolinium doped ceria (CGO). Conversely the electrodes must be porous and can be either a ceramic or cermet. Typical materials are Ni-YSZ (anode), Ni-CGO (anode), LSM (La, Sr, Mn oxide)-YSZ (cathode), LSCF (La, Sr, Co, Fe oxide, cathode).

SOFCs, especially planar cells are usually fabricated by sintering screen-printed or tape-cast ceramic powder layer(s), to allow ease of mass production and therefore reduction in cost. However, due to the multi-layer nature of the cells, sintering of the components is unavoidably under a condition that the in-plane dimensional change is constrained to a degree that depends on the exact fabrication strategy. The constraint prohibits or restrains the shrinkage within the 2-D plane of the layer or film and a bi-axial in-plane tensile stress is induced. Sintering under these conditions can lead to inadequate film density and an unfavourable pore formation [3]. Under certain circumstances, constrained sintering can lead to distortion [4-6], cracking [5-7] and even delamination [8] of cell components.

Crack formation in SOFC components is detrimental to the cell performance. For example, transverse (through) cracking in the electrolyte would lead to gas leakage. Interface cracking can lead to delamination which results in loss of electric contact and mechanical instability. From the materials processing point of view, the distortion, cracking and delamination of the cell components might be avoidable by optimising the processing procedure [4-8]. However, all these optimisation efforts have been based on a trial and error approach. A good understanding of large defect formation in constrained sintering is still lacking.

Two dimensional mechanics of cracks in sintering thin films has been analysed by Jagota and Hui [5, 8]. The predicted conditions for crack growth in sintering thin films are related to the friction of the film on the substrate and relative neck growth rate [9]. However the continuum mechanic analysis has difficulty in explaining how crack initiates. In recent years, some progress has been made in the fundamental understanding of crack initiation by using discrete element modelling (DEM) [10-13]. For example, Henrich et al [12] found that crack initiation in the constrained film is related to particle rearrangement: if rearrangement is suppressed, cracks tend to form more easily because of increased local stresses. Rasp et al [13] found that particle rearrangement also affects the delamination of ceramic strips: an easier particle rearrangement inhibiting delamination at the edges of the ceramic strips. The work of Martin et al. [10, 14] led to the conclusion that, although geometrical constraint is necessary for a defect to grow into a crack, the presence of an initial defect is not a necessary condition to initiate cracks. Nevertheless, there is still a lack of direct experimental evidence concerning how cracking is initiated.

In the current study, various ceramic films were fabricated from ceramic powders by screen printing or tape casting followed by constrained sintering. The microstructure of the sintered films was examined by high resolution SEM. The mechanism of crack formation was studied by comparing different materials and different processing strategies, and by deliberately introducing cracks in green films by indentation.

## 2 Experimental Details

Various electrolyte and electrode films were investigated in this study. In real SOFC designs, the structure can be electrode-supported, electrolyte-supported or inert supported (metal or ceramic support). As well as dimensional changes due to sintering, in actual systems there is always additionally a thermal expansion mismatch between materials, which complicates the interpretation of defect formation and evolution. To avoid the issue of thermal expansion mismatch, the substrates in this study were chosen in most experiments to have the same thermal expansion coefficient as that of the film. For example, YSZ substrates were used to support the YSZ films, fully dense CGO pellets were used to support the CGO films and so on. This ensures that any cracks observed are not the result of thermal expansion differences between film and substrate. Unless mentioned otherwise, the green ceramic films were applied to the substrates using screen printing (165 mesh screen and 2.5 mm gap) and the films were oven-dried at 120 °C. Before sintering, the dried films were heat-treated at 800 ºC in air for 10 hrs so that the binder was completely removed (i.e., the film was de-bindered). To introduce known defects in the green films, indentation was made on either dried films or de-bindered films. Sintering of the films was carried out at various temperatures and for various lengths of time using a heating rate of 3ºC/min and then cooled to room temperature at a rate of 5ºC/min. The sintered films were 10-20μm in thickness.

The 8YSZ powder used to prepare the screen-printing inks for the films was supplied by MEL Chemicals, UK. The particle size distribution of this powder was given elsewhere [15]. 8YSZ films were deposited on the substrates in the as-received condition. The substrates were commercial 3YSZ plates (Kerafol GmbH, Germany), with a thickness of 150 μm. 3YSZ has a thermal expansion coefficient which is almost the same as that of 8YSZ. Also 3YSZ is one of the toughest ceramics which allows the indentation to be made on the green film without causing damage to the thin substrate.

The LSCF ($La_{0.6}Sr_{0.4}Co_{0.2}Fe_{0.8}O_{3-\delta}$) and CGO ($Ce_{0.9}Gd_{0.1}O_{2-\delta}$) powders used for preparation of dense substrates and screen-printing inks were provided by Fuel Cell Materials (Ohio, USA), for which the specifications are shown in Table 1.

*Table 1 Manufacturer's specifications for LSCF and CGO powders*

| Powder | Surface Area ($m^2$/g) | $D_{50}$ (μm) range |
|---|---|---|
| LSCF | 6.3 | 0.3-0.6 |
| CGO | 6.0 | 0.3-0.5 |

The dense LSCF and CGO substrates (pellets)

were prepared by uniaxial pressing the powders at 150MPa for 10s, followed by isostatic pressing at a pressure of 300MPa for 30s. The green pellets of LSCF and CGO were sintered in air at high temperatures (1200°C for LSCF pellets and 1400°C for CGO pellets) for 4 hrs with a heating and cooling rate of 5 °C/min.

In the preparation of the screen-printing inks, the powders were mixed with the ink vehicle provided by Fuel Cell Materials (Ohio, USA) at a volume fraction ratio of 74%:26% and then ball milled using 2.5 mm zirconia balls for 20 hrs.

All the indentation tests were carried out at room temperature using a nano-indenter (Micromaterials, Wrexham, UK) and a micro-indenter (Leitz, Germany). In the nano indentation experiments, a Berkovich indenter tip was used. The loading and unloading rate was set to be 1 mN/s. A dwell time of 5 s at the highest load was used for all tests. In the micro-indentation experiments, a Vickers indenter was used. The load used for micro-indention was ranging from 49 mN to 981 mN. The dwell time at the set load was 10 s.

The microstructures near the indentation marks, both in green and sintered films, were examined by scanning electron microscopy (Leo 1525 and JEOL 5610). The microstructure underneath the surface (cross sectional microstructure) was examined by using FIB slice and view (Helios, FEI).

### 3 Results and discussion

3.1 Sintering and microstructures of different ceramic films

3.1.1 CGO film

The microstructure of a CGO film after constrained sintering is shown in Fig.1a and b (sintered at 1200 ºC for 1h and 1380 ºC for 6 h respectively). The CGO film was quite porous after sintering at 1200 ºC for 1 hr and did not show much further densification even after a sintering at 1380 ºC for a prolonged time (6 hrs), although the microstructure became noticeably coarser. Dilatometry (Fig.1c) showed that in free sintering the same CGO powder achieved a density more than 95 % at 1200 ºC irrespective of whether a slow ( 5 ºC/min) or a fast (20 ºC/min) heating rate was used. In contrast, the density of CGO film sintered at 1380 ºC for 6hrs is only about 85% according to image analysis. The densification of the film is thus very severely retarded by the constraint imposed by the fully dense CGO substrate. However, the microstructure of the CGO film is quite homogeneous with pores evenly distributed and equi-axed. Furthermore, no crack formation was seen in the CGO films.

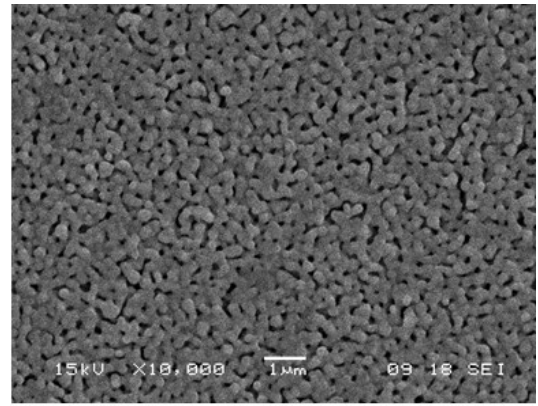

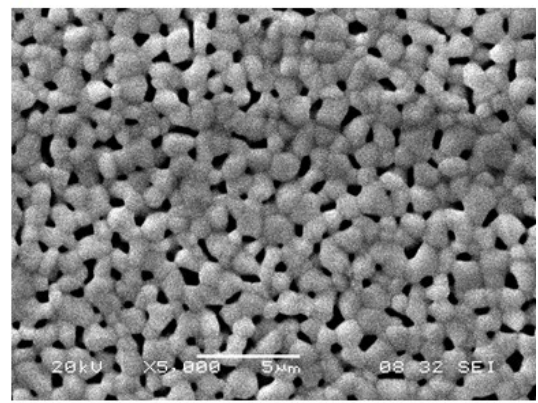

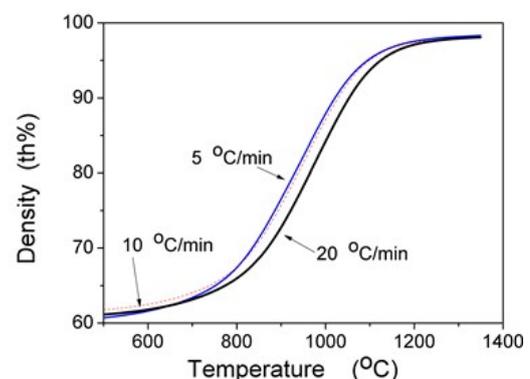

Fig. 1 SEM micrographs of the top surface of CGO films after constrained sintering at a) 1200 ºC for 1h and b) 1380 ºC for 6 h.

3.1.2 8YSZ film

The microstructure of 8YSZ films which were sintered at 1300 ºC for 1hr and 10 hrs are shown in Fig.2a and b. The sintering and microstructure of 8YSZ films have been studied systematically and published elsewhere [15]. In summary, the sintering of 8YSZ is retarded in constrained sintering more than predicted by isotropic viscous sintering model due to the anisotropic nature of the microstructure and enhanced grain growth (for a given density). The pores in 8YSZ films are elongated and preferentially oriented perpendicular to the film plane. The microstructure of 8YSZ films is heterogeneous with large variation in pore size. Nevertheless, as with the CGO films, no crack formation was seen in the 8YSZ films.

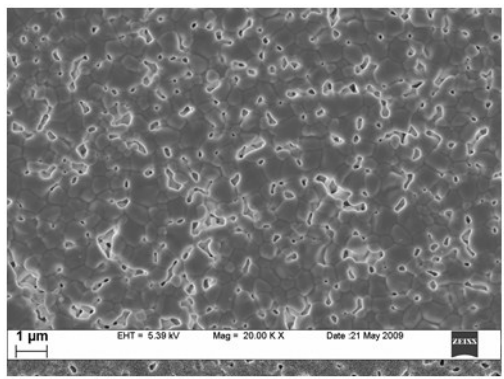

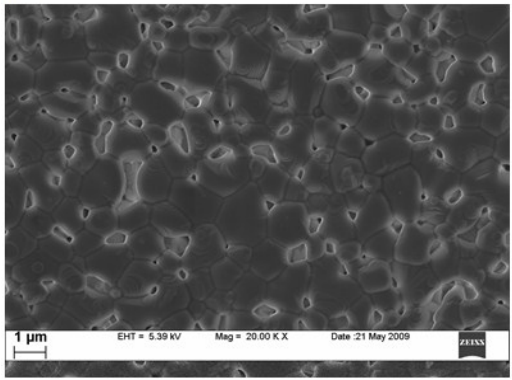

*Fig.2 SEM micrographs of the top surface of 8YSZ films after constrained sintering at 1350 ºC for a) 1 h and b) 10 h.*

3.1.3 LSCF film
Unlike the CGO and 8YSZ films, the LSCF films exhibited extensive networks of transverse cracks (Fig.3a) of the type often referred to as "mud cracks". In order to reveal the depth of the cracks, FIB milling was employed to create a local cross section (Fig.3b) which runs across one of the cracks. As shown in Fig.3c, the cracks penetrate most of the thickness of the film.

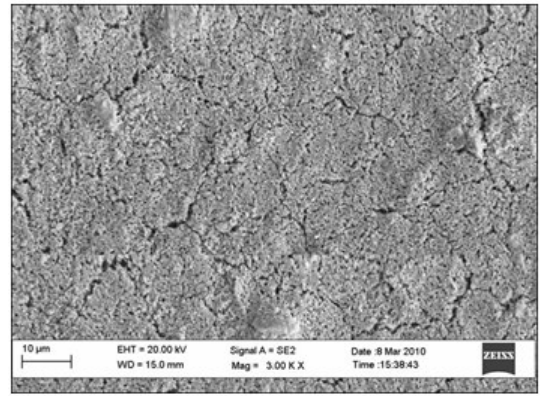

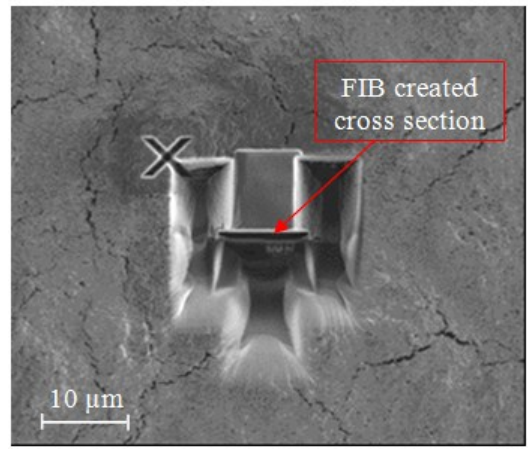

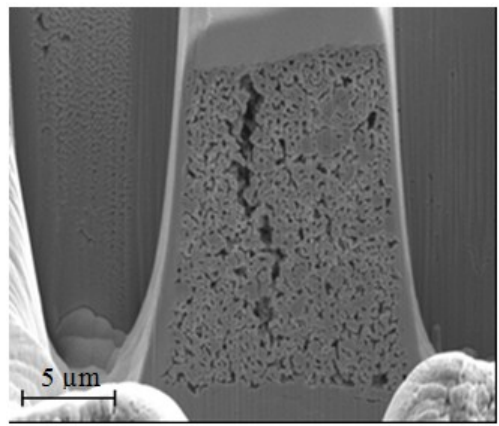

*Fig. 3 The microstructure of LSCF films after constrained sintering: a) top surface after 1000 ºC for 1 h; b) FIB milling trenches after 1100 ºC for 1h; and c) the cross sectional microstructure revealed by FIB milling.*

The cracking of the LSCF films was observed even when the sintering temperature was as low as 800 ºC and the crack opening width increased with the sintering temperature and dwell time.

As discussed later, such cracks, once formed, will never be healed in constrained sintering. Densification does occur within the 'islands' for which the constraint from the substrate is largely relieved by the crack formation, but densification of the film as a whole is prevented by the cracks. For this reason, we concentrate initially on cracking in the LSCF films.

In constrained sintering of a ceramic film, an equi-biaxial tensile stress, $\sigma$, is generated in the plane of the film due to the mismatch in unconstrained strain rate between the densifying film and the fully dense substrate. In the approximation of linear visco-elastic deformation the stress is given by [16-18]:

$$\sigma = \frac{E_{p,f} \dot{\varepsilon}_f}{1-\nu_{p,f}} \quad (1)$$

where $\dot{\varepsilon}_f$ is the unconstrained strain rate of the film and $E_{p,f}$ and $\nu_{p,f}$ are the viscosity and viscous Poisson ratio of the film.

Therefore, the sintering-induced stress is directly proportional to the strain rate and should be reduced if the strain rate is reduced. In our earlier work on 3YSZ and 8YSZ films we measured the induced stress and found that it was significantly reduced at lower temperatures when the sintering rate was lower, even though the viscosity was higher at the lower temperatures [4]. Therefore we explored different slower heating rates for the LSCF films to see if this would prevent the cracking shown in Fig.3a-c.

Fig.4 shows the densification curves for free sintering of the LSCF powder. The sintering shrinkage starts between 800 and 850 ºC, and therefore we investigated very slow heating rates for sintering the film between 700 and 900 ºC in order to maintain very small strain rates and minimise the stress. Fig.5 compares two examples of the microstructure of LSCF films sintered by using such slow heating rates. Fig.5a shows the microstructure of the film obtained by sintering using a heating rate of 5 ºC/h from 700 to 900 ºC and then cooled to room temperature. Fig, 5b shows a film sintered using a heating rate of 3 ºC/h from 700 to 800 ºC and then held at 800 ºC for 4 hrs before cooling. In both cases the slow sintering rates failed to prevent the cracking, which suggests that cracking in these films might not have been initiated during sintering.

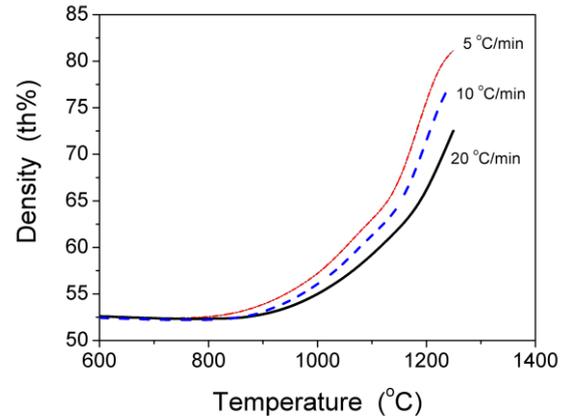

*Fig.4 Unconstrained densification of LSCF at different heating rates.*

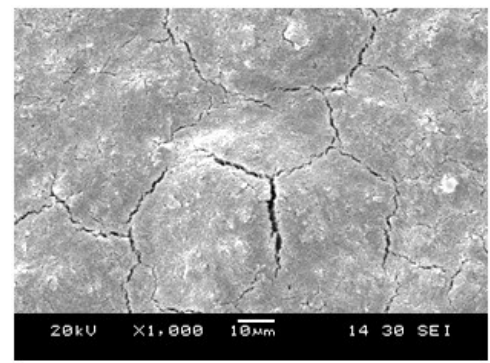

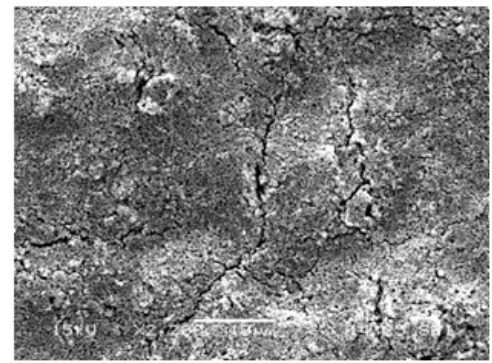

*Fig.5 The microstructure of LSCF films sintered using: a) a heating rate of 5 ºC/h from 700 to 900 ºC and no dwell; b) a heating rate of 3 ºC/h from 700 to 800 ºC and dwell at 800 ºC for 4 h.*

The LSCF film in the current case is not unique in terms of its cracking behaviour. Another example of cracking in a cathode is the LSC/CGO composite film in Fig. 6, which contains 50wt% LSC and 50wt% CGO and exhibits even more severe cracking.

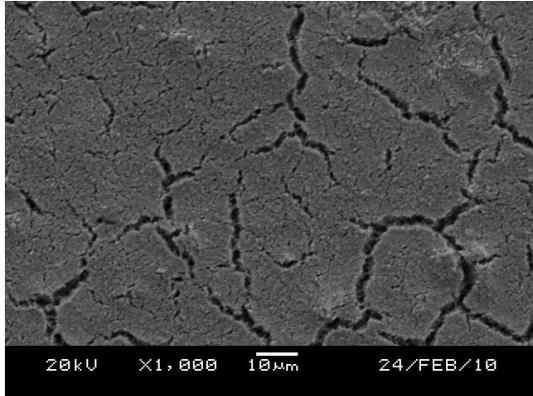

*Fig.6 SEM micrograph of LSC/CGO composite film on 3YSZ substrate sintered at 1150 ºC for 1 h.*

3.2 Cracks in green films and their evolution during sintering

Since cracks in the LSCF sintered film could have already been present in the green film before sintering, it would be useful if we know how cracks in a green film would evolve during constrained sintering.  In order to avoid complications due to the extensive cracking seen in the LSCF films, this part of the study was conducted on the 8YSZ films because they did not show such networks of cracks. Direct observation of cracks in the green film is however very difficult because the opening of the cracks could be very small and of a similar size to the particle size. Therefore, we have used indentation to introduce cracks into the green film then track the evolution of the crack in subsequent sintering.  In this way, the cracks introduced by indentation are predictable even if they are sometimes difficult to observe directly.

Fig.7a shows the cracks introduced by Vickers indentation into a dried 8YSZ film with no subsequent sintering. Some "concentric" cracks are visible by SEM in this case along the outer edges of the indentation mark. There are no major radial cracks visible at the four sharp corners of the indentation mark  Fig.7b shows that after early stage sintering at 1150 ºC for 1h, the opening of the "concentric" cracks increased and other cracks (not visible in the dried film) are now evident along the contacts with the indenter edges.

Fig.8a shows the cracks introduced by indentation of an 8YSZ film after binder burn-out at 800 ºC for 10 h. In contrast to the case of the dried film, in debindered film there are radial cracks visible at the corners of the indent, but no concentric cracks are evident. The reason for this difference in cracking behaviour is that the dried film and the debindered film have very different mechanical properties [19].

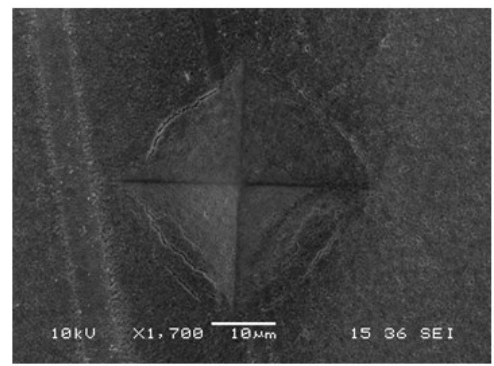

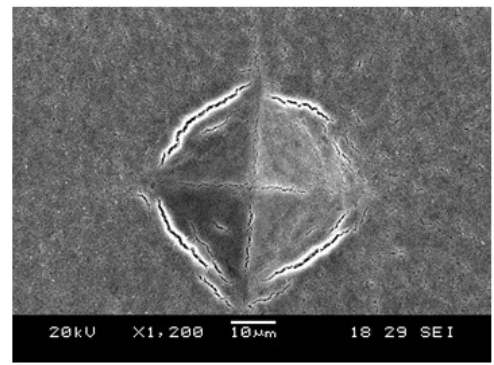

*Figs.7 Cracks introduced by indentation of a dried 8YSZ film: a) in the dried state; and b) after subsequent sintering at 1150 ºC for 1h.*

Based on microindentation experiments, the hardness of the debindered film (0.33GPa) was about 2.5 times of that of the dried film (0.13 GPa). The bonding after binder burn-out is at direct contacts between ceramic particles formed by some small degree of sintering and therefore the debindered film has a 'brittle' nature. In contrast the dried film still has polymeric binder between solid particles and does not have strong inter-particle bonds. It therefore behaves more like a partly plastic material.  We speculate that the "concentric" cracks in the dried film could be due to the 'tearing' of the interparticle polymer driven to increase their opening after sintering (Fig.8b and c) and the extent of opening increased with both temperature and dwell time, as can be seen by comparing Fig.8b with Fig.8c. This behaviour is consistent with the observed effect of sintering on crack opening in the LSCF films described earlier. Although the crack opening increased during sintering, the length of the cracks was observed to be unchanged. Fig.9a gives the length of the radial crack and indentation mark size (half diagonal length) generated by a Vickers indenter on a debindered film as a function of load. When the load was sufficiently small ( <100mN), no radial cracks were observed after sintering. This could be because the cracks formed in the green film were either small enough to be healed during subsequent sintering, or they were not formed in the first place. For the loads larger than 200mN, there is a well defined relationship between indentation load and the crack length.

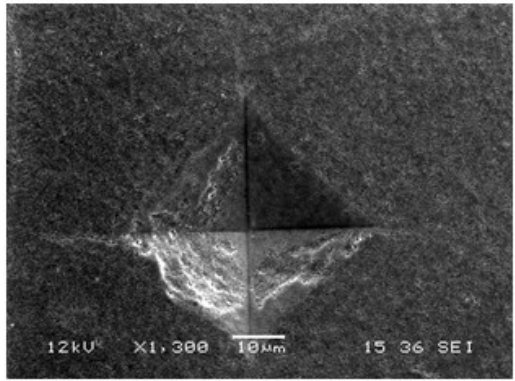

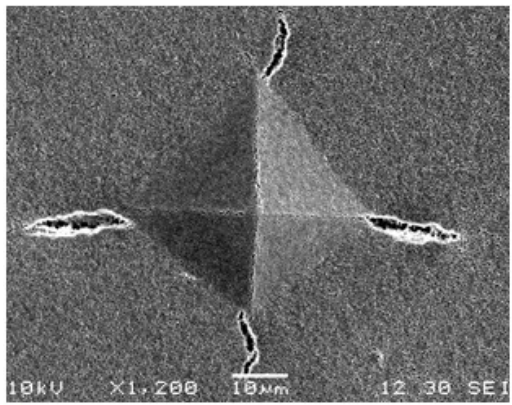

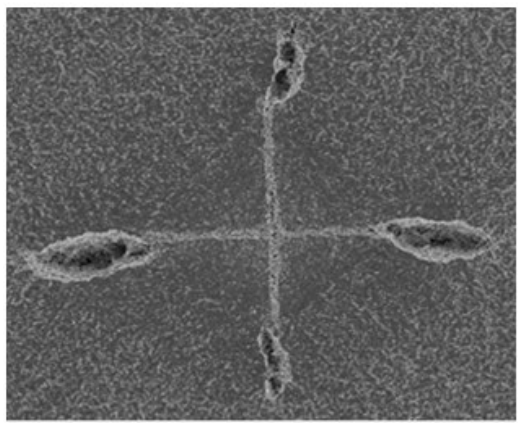

*Fig.8 Cracks introduced by indentation after binder burn-out of an 8YSZ film: a) after indentation; b) after subsequent sintering at 1250 ºC for 1h: and c) after a further sintering at 1350 ºC for 1h.*

ceramic material such as the debindered film, crack formation at the sharp corners of the Vickers indenter is expected. Those cracks were observed

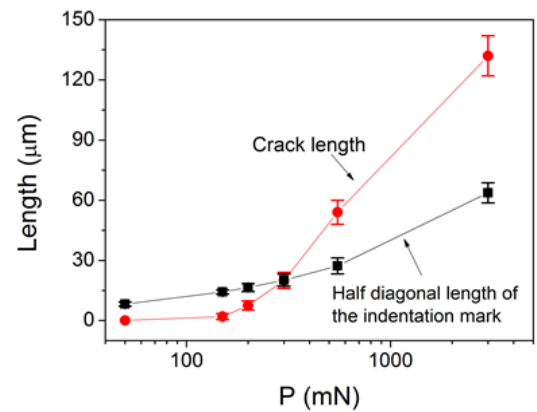

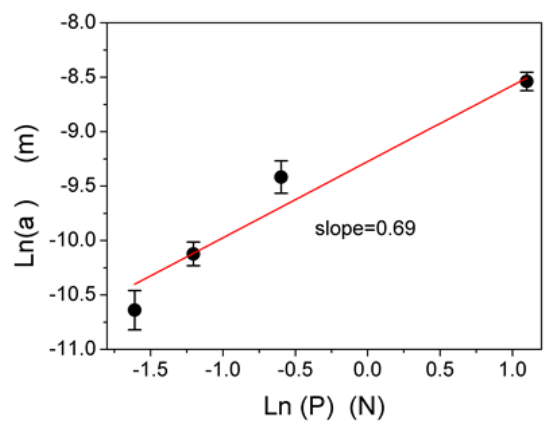

*Fig.9 a) Radial crack length and indentation mark size (half diagonal length) generated by a Vickers*

*indenter with different loads on an 8YSZ film, b) Ln(a) vs. Ln(P) plot.*

According to indentation microfracture theory for a brittle material, the crack length ($a$) is directly proportional to $P^{2/3}$ [20]. Here $a$ is measured from the centre of the indentation mark, (equal to the sum of the radial crack length and the half diagonal length of the indentation mark) and $P$ indentation load. Therefore, if Ln($a$) is plotted against Ln($P$), a linear relationship with a slope equal to 2/3 should be obtained. As shown in Fig.9b, the linearity between Ln(a) and Ln(P) appears to be reasonable and the slope is very close to 2/3. This is consistent with the cracks being formed in the green film and not growing in length during sintering.

In Fig.10 indentation marks (after sintering) made on a dried CGO film and a debindered CGO film are shown. The cracking pattern in the debindered CGO film (Fig. 10b) is very similar to that seen in the 8YSZ film (Fig. 8 b), but the dried CGO film (Fig.10a) shows a different cracking pattern from that of 8YSZ (Fig.7b) in that the circumferencial cracks are absent. This might be due to lower friction between CGO particles and the diamond indenter. Nevertheless, the crack lengths in the CGO films were observed to be unchanged during sintering, as was observed for the 8YSZ films.

Here the absence of crack length increase in YSZ or CGO films does not necessarily contradict to the observation of Bordia and Jagota [9] who noticed some degree of crack propagation in both glass and alumina films during film sintering. Firstly the crack extentions observed in [9] were all small as compared to the cracks. Secondly, as pointed out in [9] the crack propogation criteria is dependent not only on the interaction between substrate and the film (friction), but also on the ratio of the rate of neck growth due to sintering to the linear shrinkage rate during sintering. So the different films are expected to have different mechanical behaviour during constrained sintering.

3.3 Healing of cracks in co-sintering

In an early work, a variational principle sintering model predicts that in constrained sintering there exists a critical pore size above which the pore grows instead of shrinks [15]. In this work we found that for all the different ceramic films, when the cracks present in the green film before sintering were relatively large, they could not be eliminated in the subsequent sintering, instead the crack opening increased. This is in consistent with the variational principle sintering model, if the cracks are regarded as narrow pores. Regarding the critical size of crack, all the radial cracks (after constrained sintering) introduced in debindered 8YSZ films by using different loadings were measured on SEM images. The smallest crack was approximately 2μm in length. Therefore the critical crack size can be estimated to be equal to or smaller than 2μm for 8YSZ film.

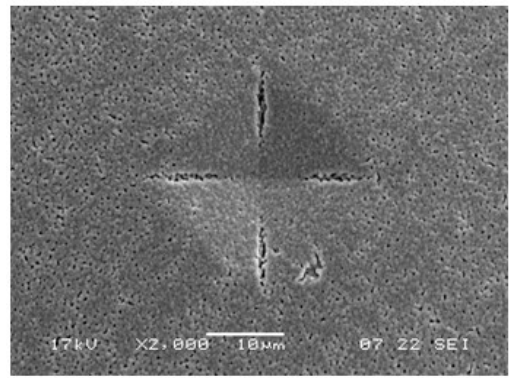

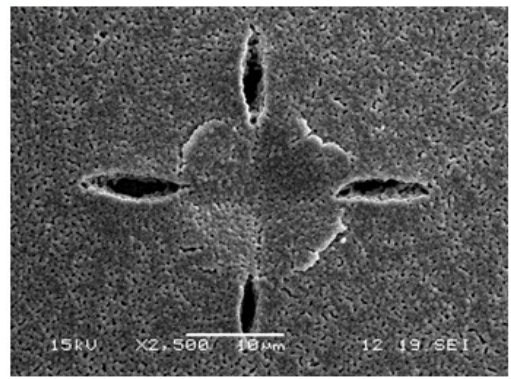

*Fig.10 The microstructure of a) indentation on dried CGO film sintered at at 1250 ºC for 4 hrs, b) indentation on debindered CGO film sintered at 1250 ºC for 4 hrs.*

Given that many SOFC cells are made by co-sintering, some experiments were carried out to explore how cracks in a green film would evolve in co-sintering, where the constraint is smaller than when fully constrained in two dimensions. An 8YSZ film was therefore tape-cast on a partially sintered 8YSZ substrate. The substrate was sintered at 1150 ºC for 10 min to achieve a similar density (about 60%) as the tape-cast green film. After binder burn-out of the green film at 800 ºC for 10 hrs, indentation was made on top of the green film. Fig.11a shows the cracks formed at the sharp corners of the indentation mark after sintering at 1250 ºC for 1h. The fact that cracks in the green film still widen during co-sintering indicates that the densification rate of the film was faster than that of the substrate at this relatively early stage of sintering. Although both substrate and green film have similar density to start with, the packing of the tape-cast film is possibly better and more homogeneous, which would make the initial unconstrained densification rate of the film faster than that of the substrate. Fig.11b shows these cracks were all closed after further sintering at a higher temperature (1350 ºC for 1h). To examine the depth of the cracks, FIB milling was empoloyed to create local cross sections at the locations indicated in Figs. 11a and b and these are shown in Figs. 11 c and d respectively.

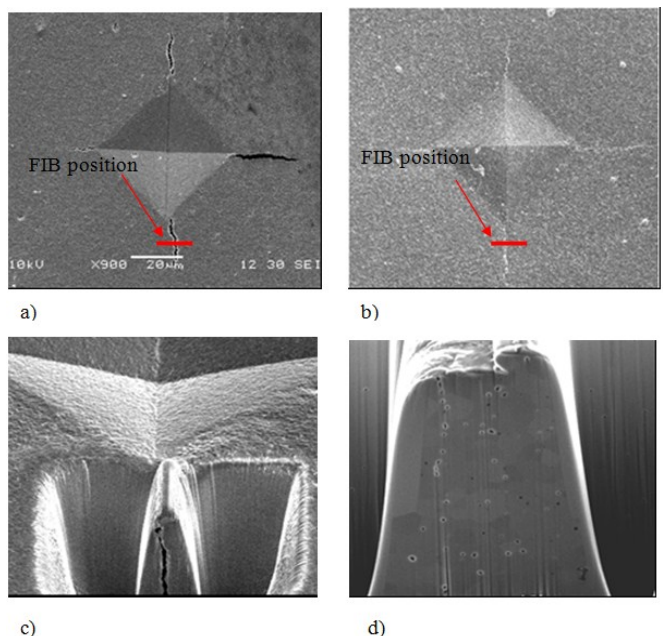

*Fig11 Cracks introduced by indentation in green films (debindered) supported on partially sintered substrate, a) after sintering at 1250 ˚C for 1 hr, b) after sintering at 1350 ˚C for 1hr, c) the cross section created by FIB corresponding to Fig.11a, d) the cross section view corresponding to Fig.11b.*

The crack after co-sintering at 1250 ºC for 1 h was observed to penetrate vertically into the film (Fig.11c). However, after further sintering no visible crack can be seen on the cross section and the cracks had healed, although the marks of the original cracks still can be seen from the top surface as shown in Fig.11b. It is speculated that the cracks were healed as a result of a compressive stress generated by the unconstrained sintering rate of the substrate being faster than that of the film in the later stages of sintering.

3.4 Avoidance of cracks in constrained sintering LSCF films

The experiments described above on electrolyte films have shown that cracks in the green films cannot be healed during constrained sintering. For this reason the preparation of crack-free green LSCF films is essential for fabricating high quality sintered LSCF films. The interactions between the liquid phase, the organic additives (such as binder) and solid particles in the ink are critical in controlling the mechanical behaviour of the green film during drying and binder burn-out. In our work, the properties of the LSCF ink were modified by dilution with terpineol (Sigma Chemical Co. USA) followed further by ball-milling for 12 hrs. A green LSCF film was prepared using the modified LSCF ink and sintered at 900° C for 4 hrs. As shown in Fig. 12, the sintered film showed no cracks. This indicates the crack formation in constrained sintered films can be avoided by preparation of 'crack-free' green films.

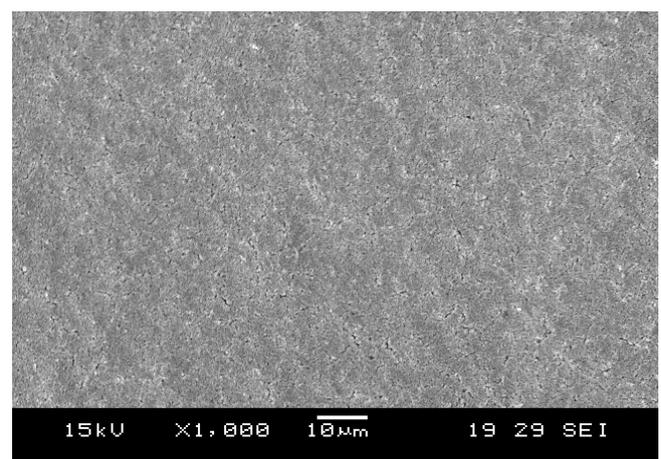

*Fig.12, LSCF film prepared from the modified ink and sintered at 900°C for 4 hrs.*

## 4 Summary

Films of 8YSZ, CGO and LSCF were prepared on dense substrates (which had the same thermal expansion coefficient as the supported film) using screen printing or tape-casting followed by drying, binder burn-out and sintering. The densification of the films was observed to be retarded by the constraint from the substrates in agreement with other reported studies. The microstructure of the sintered films was observed to be qualitatively different for the different materials. While cathode films (LSCF and LSC/CGO composite) presented severe mud-cracking, YSZ and CGO films showed no mud-cracking. Furthermore, while YSZ film presented some local crack-like defects or large pores, CGO film presented a quite homogeneous porous microstructure and did not reach high density.

To understand how cracks forms and evolve during constrained sintering, artificial cracks were introduced in the green ceramic films of 8YSZ and CGO using indentation. It was found that when the crack length in green films was larger than a critical size, constrained sintering always led to the widening of the cracks, while the crack length was unchanged. In contrast, in co-sintering a crack originally introduced by indentation of the green film was first observed to widen in early stage sintering, but then heal at a later stage.

The initiation of cracking in LSCF films was studied by varying the heating schedule at the earliest stage of sintering. It was found that cracking always occurred no matter how slow the heating rate (or strain rate) was at this earliest stage. However, it was possible to avoid cracking by modifying the ink rheology. These observations indicate that the cracks observed after sintering probably originated from a crack network in the green film which was not observable because the crack openings were too small. This suggests a general principle: namely, that cracks in sintered films can be avoided by preparation of crack-free green films.

## Acknowledgements

This research was carried out as part of the UK Supergen consortium project on "Fuel Cells: Powering a Greener Future". The Energy Programme is an RCUK cross-council initiative led by EPSRC and contributed to by ESRC, NERC, BBSRC and STFC.